\documentclass[numberedappendix,apj,twocolumn]{emulateapj}

\newcommand{\zy}{$z_{850} - Y_{105}$}
\newcommand{\yj}{$Y_{105} - J_{125}$}
\newcommand{\jh}{$J_{125} - H_{160}$}

\newcommand{\iFilter}{$i_{775}$}
\newcommand{\zFilter}{$z_{850}$}
\newcommand{\yFilter}{$Y_{105}$}
\newcommand{\jFilter}{$J_{125}$}

\newcommand{\hFilter}{$H_{160}$}

\newcommand{\msol}{$M_\odot$}

\shorttitle{Structure and Morphologies out to $z\sim7-8$}
\shortauthors{Oesch et al.}

\begin{document}

\title{Structure and Morphologies of $z\sim7-8$ Galaxies from ultra-deep WFC3/IR Imaging of the HUDF
\altaffilmark{1}}

\altaffiltext{1}{Based on data obtained with the \textit{Hubble Space Telescope} operated by AURA, Inc. for NASA under contract NAS5-26555. }

\author{P.A. Oesch\altaffilmark{2},
R.J. Bouwens\altaffilmark{3}, 
C.M. Carollo\altaffilmark{2}, 
G.D. Illingworth\altaffilmark{3}, 
M. Trenti\altaffilmark{4}, \\
M. Stiavelli\altaffilmark{5},
D. Magee\altaffilmark{3}, 
I. Labb\'{e}\altaffilmark{6}, 
M. Franx\altaffilmark{7}
}

\altaffiltext{2}{Institute for Astronomy, ETH Zurich, 8092 Zurich, Switzerland; poesch@phys.ethz.ch}
\altaffiltext{3}{UCO/Lick Observatory, University of California, Santa Cruz, CA 95064}
\altaffiltext{4}{University of Colorado, Center for Astrophysics and Space Astronomy,
389-UCB, Boulder, CO 80309, USA}
\altaffiltext{5}{Space Telescope Science Institute, Baltimore, MD 21218, United States}
\altaffiltext{6}{Carnegie Observatories, Pasadena, CA 91101, Hubble Fellow}
\altaffiltext{7}{Leiden Observatory, Leiden University, NL-2300 RA Leiden, Netherlands}


\begin{abstract}
We present a first morphological study of $z\sim7-8$ Lyman Break galaxies (LBGs) from \citet{oesch09b} and \citet{bouw09b} detected in ultra-deep near-infrared imaging of the Hubble Ultra Deep field (HUDF) by the HUDF09 program. With an average intrinsic size of $0.7\pm0.3$ kpc these galaxies are found to be extremely compact having an average observed surface brightness of $\mu_J\simeq26$ mag arcsec$^{-2}$, and only two out of the full sample of 16 $z\sim7$ galaxies show extended features with resolved double cores.
By comparison to lower redshift LBGs it is found that only little size evolution takes place from $z\sim7$ to $z\sim6$, while galaxies between $z\sim4-5$ show more extended wings in their apparent profiles. 
The average size scales as $(1+z)^{-m}$ with $m=1.12\pm0.17$ for galaxies with luminosities in the range (0.3-1)$L^*_{z=3}$ and with $m=1.32\pm0.52$ for (0.12-0.3)$L^*_{z=3}$, consistent with galaxies having constant comoving sizes. The peak of the size distribution changes only slowly from $z\sim7$ to $z\sim4$. However, a tail of larger galaxies ($\gtrsim 1.2$ kpc) is gradually built up towards later cosmic times, possibly via hierarchical build-up or via enhanced accretion of cold gas. 
Additionally, the average star-formation surface density of LBGs with luminosities (0.3-1)$L^*_{z=3}$ is nearly constant at $\Sigma_{SFR}=1.9$ \msol yr$^{-1}$kpc$^{-2}$ over the entire redshift range $z\sim4-7$ suggesting similar star-formation efficiencies at these early epochs.
The above evolutionary trends seem to hold out to $z\sim8$ though the sample is still small and possibly incomplete.

\end{abstract}

\keywords{galaxies: evolution ---  galaxies: high-redshift --- galaxies: structure}

\section{Introduction}
The newly installed WFC3/IR camera on the Hubble Space Telescope has opened up a new territory in the study of galaxies at $z\gtrsim6.5$. Its increased capability has lead to numerous detections of galaxies at $z\gtrsim6.5$ already in the first epoch data of the HUDF09 \citep{oesch09b,bouw09b,mclure09,bunker09}.

Understanding the evolution of galaxy sizes and morphologies out to $z\gtrsim6.5$ can provide essential clues to galaxy formation models. While hydrodynamical galaxy formation simulations of LBGs \citep[e.g.][]{finlator06,night06,nagamine08} have focussed on the prediction of the evolution of the galaxy luminosity and mass functions, the sizes of these galaxies have not been addressed in great detail so far. 
By providing these results we hope to stimulate interest in this key aspect of galaxies at early times.

Based on semi-analytical modeling, the sizes of Lyman Break Galaxies (LBGs) at fixed luminosity are expected to slowly decrease towards earlier cosmic times \citep{mmw98}, which is in good agreement with earlier observations of LBGs \citep[e.g.][]{ferg04,bouw04}.

The UV morphologies of LBGs have been studied between $z\sim2-6$ from both ground-based and Hubble Space Telescope imaging. They have been found to be very compact, but often containing multiple cores, especially at brighter magnitudes \citep[e.g.][]{giavalisco96,lowenthal97,ravindranath06,lotz06,law07,conselice09}. The dominant mechanism for star-formation in these galaxies is still debated. LBGs with multiple cores have been interpreted as merging systems, with star-formation triggered by this interaction, which is a natural scenario in a hierarchical universe \citep[e.g.][]{somerville01,overzier08}. However, these individual cores could also originate from individual star-forming clumps within a larger gas-dominated disk galaxy, whereby this clumpy state is kept alive due to gas streams replenishing the disk galaxy with cold gas \citep[e.g.][]{dekel09}.

Here we present the highest resolution observations available to date of galaxies from a time when the universe was only $\sim800$ Myr old.
Our analysis is primarily based on the sample of 16 $z\sim7$ LBGs from \citet{oesch09b}. We also discuss the 5 $z\sim8$ objects identified by \citet{bouw09b}.
These galaxies are detected in the first epoch ultra-deep WFC3/IR imaging of the HUDF09 survey, which reaches $\sim29$ mag ($5\sigma$) in \yFilter\jFilter\hFilter photometry over an area of 4.7 arcmin$^2$ covering the Hubble ultra-deep field \citep[HUDF;][]{beck06}.
 The 16 $z\sim7$ galaxy candidates are selected based on their (\zy) vs. (\yj) colors using the Lyman Break Technique \citep[e.g.][]{steidel96} and have an expected redshift distribution $z\sim6.5-7.5$ with a median at $\langle z\rangle=6.8$. Similarly, the $z\sim8$ sample is selected based on their (\yj) vs. (\jh) colors. For more information on the survey and the galaxy sample we refer to \citet{oesch09b} and \citet{bouw09b}.
We also include galaxies from $z\sim4-6$ identified as dropout galaxies in the optical HUDF data for comparison with the $z\gtrsim6.5$ population.

We adopt $\Omega_M=0.3,\Omega_\Lambda=0.7,H_0=70$kms$^{-1}$Mpc$^{-1}$, i.e. $h=0.7$. Magnitudes are given in the AB system \citep{okeg83}. 
We express galaxy UV luminosities in units of the characteristic luminosity at $z\sim3$ being $M_{1600}(z=3)=-21.0$ \citep{steidel99}.

\begin{figure}[tbp]
	\centering
		\includegraphics[width=\linewidth]{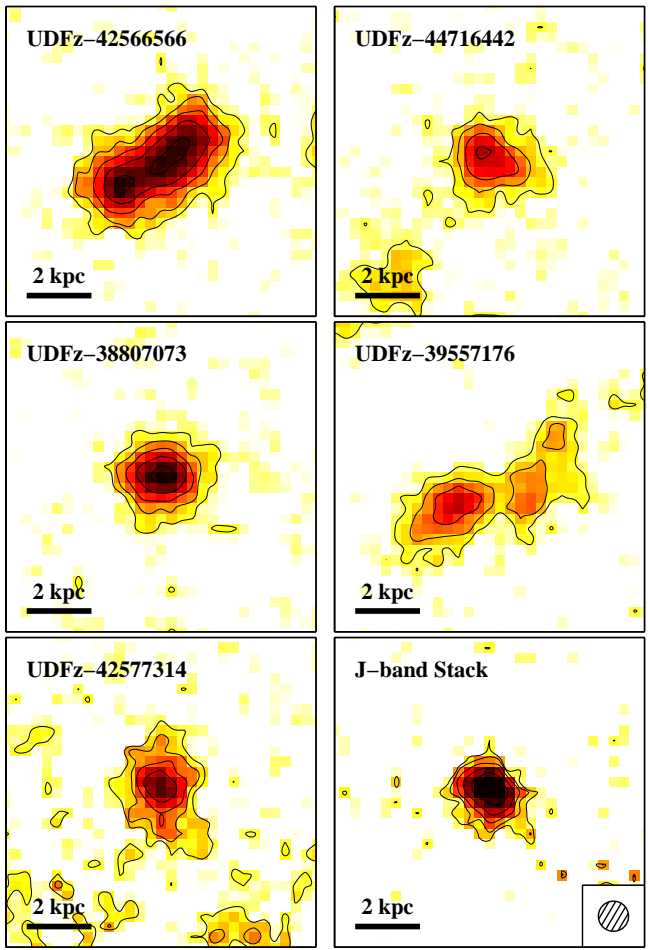}
	\caption{Surface brightness contours of the five brightest galaxies in our sample and of a \jFilter\ stack of the remaining 11 fainter galaxies (last panel in lower right). The first five images are superpositions of \yFilter, \jFilter, and \hFilter\ exposures, with the contour lines corresponding to $\mu_\mathrm{YJH}=23.5-25.5$ mag/arcsec$^2$ in steps of 0.5 mag/arcsec$^2$. The bar in the left corner indicates 2 kpc (physical) at $z=6.8$ the expected mean redshift of these galaxies. All images are 1\farcs8 on a side. The size (FWHM) of the \jFilter\ PSF is shown as an inset in the lower right panel for comparison.
	}
	\label{fig:contour}
\end{figure}

\section{Morphologies at $z\sim7$}

The $z\sim7$ galaxies are extremely compact as can be seen from the contour plots in Figure \ref{fig:contour}, where we show the individual summed \yFilter, \jFilter, and \hFilter\ observations of the five brightest sources and additionally a \jFilter\ stack of the remaining 11 fainter ones
\citep[for stamps of individual sources see][]{oesch09b}. 
As can be seen, the average $z\sim7$ galaxy appears to be very symmetric and compact; with two exceptions, no extended features can be identified. The two exceptions are:

a) The galaxy UDFz-42566566 is the brightest galaxy in our sample and consists of two clearly distinct components, separated by 2 kpc. These two each contribute about the same amount of light (1:1.2) with very similar colors. They have individual half-light radii of 0.5 kpc and 0.8 kpc, respectively, very similar to the compact galaxies in our sample. One interpretation for the origin of the individual components is therefore that they are in a merging phase. The linear geometry of the whole galaxy, however, may also suggest that the individual clumps are star-forming regions within a disk structure, similar to what has been found at $z\sim2$ in observations and simulations.

b) The galaxy UDFz-39557176 also consists of at least two components. The total light of this galaxy is dominated by a slightly elongated central structure, which has a fainter counterpart about 2 kpc away to the NW. The flux ratio of these two is 1:1.4. It is worth noting that this galaxy has been split into two sources in the \citet{mclure09} catalog. However, the two knots are most probably physically connected and are about to merge with each other. The fainter component to the NW shows a significantly redder \yj\ color compared to the central core by 0.2 mag, and also a redder \jh\ color by 0.1 mag. This may indicate that the second component consists of older stellar populations. However, a more speculative explanation could be that the second component is reddened due to dust from the central core. This would imply that these galaxies contain a more extended gas disk than what can be seen from their UV light.

\section{Comparison to LBGs at $z\sim4-6$}

In order to quantify the evolution of galaxy structures across cosmic time, we compare the $z\sim7$ galaxies with LBGs identified at $z\sim4-6$. We focus on three main aspects: (1) the size evolution, (2) the evolution of the average galaxy light profile, and (3) the evolution of the surface density of star-formation in these galaxies.

 \begin{figure}
	\centering
		\includegraphics[width=\linewidth]{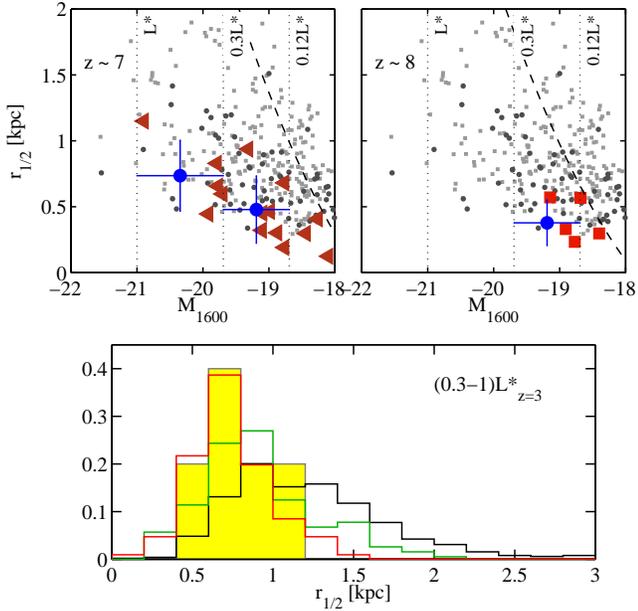}
	\caption{ \textit{Top:} Luminosity-Size relation for $z\sim5-8$ galaxies in the WFC3/IR observations. The colored points are the $z\sim7$ (left) and $z\sim8$ galaxies (right). Dark gray dots represent samples of $z\sim5-6$ based on WFC3/IR observations, while light gray squares are $z\sim4$ galaxies based on optical ACS data for comparison. All sizes are PSF corrected. The dashed line indicates the 50\% completeness for galaxies with exponential profiles shifted to the absolute magnitude at the given redshift. The vertical dotted lines correspond to the edges of the luminosity bins with $0.12, 0.3, 1 L^*_{z=3}$. The blue points mark the mean of the $z\sim7$ population in the given luminosity bin.
	\textit{Bottom:}  Comparison of the size distribution of $z\sim4-7$ galaxies in the luminosity range (0.3-1)$L^*_{z=3}$. The filled yellow histogram is for $z\sim7$ galaxies, the others are for lower redshift LBGs (red: $z\sim6$, green: $z\sim5$, black: $z\sim4$). In order to increase statistics, this plot is based on GOODS for the galaxies of $z\sim4-6$, where the sizes are measured from SExtractor in the \iFilter\ ($z\sim4$) and \zFilter ($z\sim5-6$). The $z\sim7$ measurements are based on the \jFilter. The main difference is the tail towards larger sizes at later cosmic times, while the peak of the size distribution changes little.
	}
	\label{fig:LumSizeSFR}
\end{figure}

\begin{figure}[htbp]
	\centering
		\includegraphics[width=\linewidth]{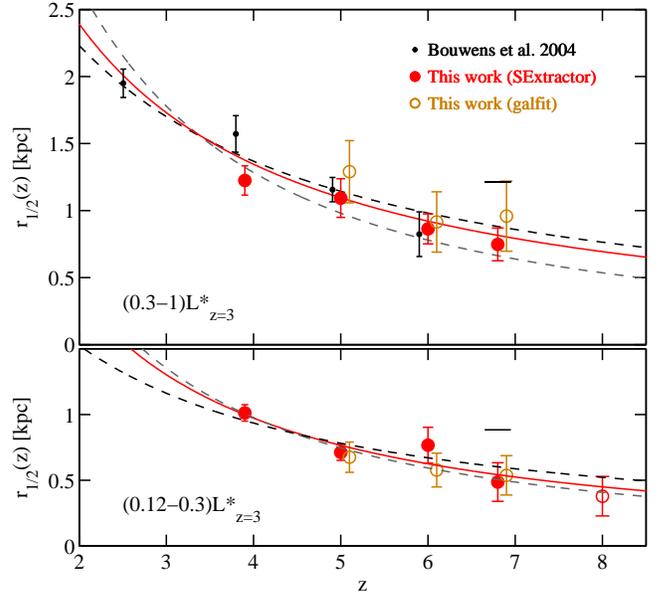}
	\caption{The evolution of the mean galaxy size across the redshift range $z\sim2-8$ in two different luminosity ranges (0.3-1)$L^*_{z=3}$.
	 (top) and (0.12-0.3)$L^*_{z=3}$ (bottom). 
	The data from this work is plotted as red dots and is based on the PSF corrected SExtractor half-light radii. Also reported are the measurements from \citet{bouw04} based on optical ACS data of the same luminosity range. 
	These points are repeated in the lower panel for comparison. 
	Mean half-light radius measurements from Sersic fits with galfit are shown as orange circles, offset from their redshift by 0.1. The dashed lines indicate the scaling expected for a fixed halo mass ($\propto H(z)^{-2/3}$; black) and at fixed halo circular velocity ($\propto H(z)^{-1}$; gray). The red lines correspond to the best-fit to the observed evolution described by $(1+z)^{-m}$, with $m=1.12\pm0.17$ for the brighter luminosity bin, and $m=1.32\pm0.52$ at fainter luminosities. These are formally identical and are consistent with the $m\sim1$ value derived previously. The short black line at $z\sim7$ indicates the mean sizes found when projecting the observed $z\sim4$ sample to $z\sim7$, while keeping their luminosities and
physical sizes constant in order to test for measurement biases due to cosmological surface brightness dimming.
	}
	\label{fig:sizeevol}
\end{figure}

\subsection{Size evolution}

Galaxy sizes are measured using circular apertures containing 50\% of the galaxies' light. We use the observed half-light radius from SExtractor, $r_{1/2,SE}^{obs}$, and correct it for PSF broadening according to $r_{1/2,SE}=\sqrt{(r_{1/2,SE}^{obs})^2-r_{PSF}^2}$. The radius $r_{PSF}$ of a point source is 0\farcs12 in the \jFilter, and 0\farcs11 in the \yFilter\ observations. 
These measurements are checked against the higher resolution optical data for the $z\sim5$ population where size measurement from both ACS (\iFilter) and WFC3/IR (\yFilter) are available. No significant differences are found, showing the robustness of these size estimates even for such faint, small galaxies and the validity of the simple PSF correction.

As a second check, we adopt size measurements based on \texttt{galfit} \citep{peng02}. Single sersic profiles are fitted to these galaxies with sersic indices fixed at $n=1.5$, the value measured for a mean stack of $z\sim4$ LBGs. The half-light radii estimates of the best-fit models are in good agreement with the SExtractor measurements with a dispersion of $\sigma_r=0\farcs05$ and no bias.
Similar results are found when using an average Sersic index  $n=1$, or $n=3$.
However, \texttt{galfit} fails to return reliable measurements when the light profile is not well approximated by a single component fit and we use the SExtractor measurements as our fiducial ones. 
When appropriate, we also comment on the implications of using the \texttt{galfit} measurements.

Due to selection effects, large, low-surface brightness galaxies will be missed in our catalog. We estimate this bias by inserting artificial galaxies of fixed profiles into the science images and rerunning the detection algorithm with the same parameters as for the creation of the original catalogs \citep[for more details see][]{oesch07}. The galaxy profiles are chosen to follow an exponential Sersic function ($n=1$) and a de Vaucouleur profile ($n=4$). 

In the upper panel of Figure \ref{fig:LumSizeSFR} we plot the half-light radii against the absolute magnitudes for the $z\sim7-8$ galaxies and both $z\sim5-6$ and $z\sim4$ samples. As can be seen, the $z\sim7$ sample is complete for all galaxies at 1 kpc down to $M_{1600}=-18.7$, corresponding to $0.12L^*_{z=3}$. At these same luminosities the galaxies in the $z\sim8$ sample are close to the completeness limit. We therefore do not include the $z\sim8$ sample when fitting scaling relations as a function of redshift in our analysis.

As can be seen from the figure, at fixed luminosities, the sizes evolve only little from $z\sim4-8$. However, the lower redshift population contains an extended tail towards galaxies with sizes $\gtrsim1.2$kpc, which is not seen at higher redshifts as shown in the lower panel of Figure \ref{fig:LumSizeSFR}. In order to increase the statistics, the $z\sim4-6$ galaxy samples are taken from GOODS based on ACS imaging \citep[e.g.][]{giav04a,giav04b}. The peaks of the distributions are between $0.8-1$kpc for all redshifts. The $z\sim4$ population contains a significant population of galaxies with $\gtrsim1.2$kpc, which are completely absent at $z\sim7$. Firm confirmation of the lack of such large objects at $z\sim7$ will require larger area surveys, but the trend seen from $z\sim6$ to lower redshift is very suggestive.
The above indicates a scenario in which these primordial star-forming galaxies are formed as small clumps with a size of $\sim0.4-0.8$kpc independent of luminosity. Towards later cosmic times hierarchical build-up or enhanced accretion of cold gas starts to produce larger galaxies.

The evolution of the average sizes of galaxies from $z\sim2-8$ are plotted in Figure \ref{fig:sizeevol} for two different luminosity ranges, (0.3-1)$L^*_{z=3}$ (top) and (0.12-0.3) $L^*_{z=3}$ (bottom). 
The expected size scaling from semi-analytical models is $H(z)^{-1}$, at fixed halo circular velocity, or $H(z)^{-2/3}$ at fixed halo mass \citep{mmw98}, where $H(z)$ is the Hubble parameter at redshift $z$ which scales as $\sim(1+z)^{3/2}$ at $z>2$.
The observed size evolution at fixed luminosity is fitted with a scaling of the form $(1+z)^{-m}$. The two fits are formally identical with $m=1.12\pm0.17$ and $m=1.32\pm0.52$, respectively.
This is in agreement with previous estimates where the sizes were found to scale roughly according to $(1+z)^{-1}$ \citep[][]{bouw04,bouw06}. However, $H(z)^{-1}$ \citep{ferg04,hathi08b} cannot be ruled out as both scalings are very flat over our redshift range and diverge only at lower $z$. 

Note that the observed evolution in sizes is not an artifact of cosmological surface brightness dimming.  We 
verified this by artificially redshifting our $z\sim4$ sample (\iFilter\ band) to $z\sim7$ (\jFilter\ band) 
and remeasuring their sizes with SExtractor. For the more luminous galaxies, the input size is recovered
perfectly, while for the lower luminosity sample only a small bias of $\sim10$\%
towards smaller sizes is found due to the flux loss in the galaxy wings.

\begin{figure}[tbp]
	\centering
		\includegraphics[width=\linewidth]{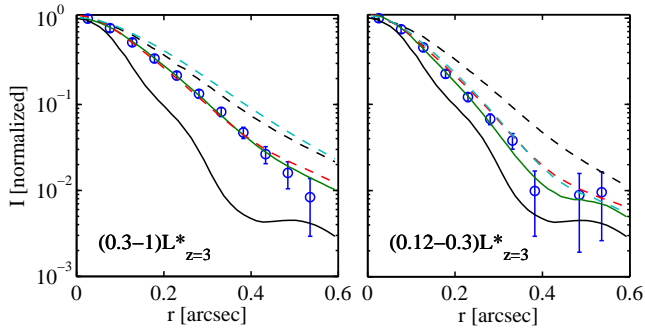}
	\caption{Apparent intensity profiles of the stacked $z\sim4-7$ galaxies in two different luminosity ranges, (0.3-1)$L^*_{z=3}$ (left) and (0.12-0.3)$L^*_{z=3}$ (right). The blue points are the data, while the black line is the PSF from stars in the \jFilter\ observation. The best-fit model of the $z\sim7$ population is shown as a solid green line. Models for lower redshift galaxies are shown as dashed lines in red ($z\sim6$), cyan ($z\sim5$) and black ($z\sim4$). For galaxies with luminosities (0.3-1)$L^*_{z=3}$, the $z\sim4-5$ population shows clearly extended wings with respect to higher redshift galaxies, while in the lower luminosity bin, the evolution is slower. Note that this is consistent with increasing physical sizes towards lower redshifts due to increasing angular diameter distances.
	}
	\label{fig:profiles}
\end{figure}

\subsection{Average Profiles}
\label{sec:avProfiles}
In order to test the uniformity of average galaxy sizes over these redshifts, we measure their average profiles based on stacking galaxies from fixed luminosity ranges and fitting these profiles with \texttt{galfit}. For the $z\sim7$ galaxies this is done in the \jFilter\ band, corresponding closely to restframe 1600 \AA. The galaxy stamps are cleaned of close neighbors using the cleaning algorithm of \texttt{ZEST+} \citep[][in preparation]{zest+}. The algorithm masks neighboring sources in the images and replaces contaminated pixels with background noise.
The point-spread function (PSF) was estimated on three stars which showed no saturation in individual exposures. We have also ensured that the central pixels of the stars have not been affected by cosmic-ray rejection in the data reduction pipeline. 

The above procedure is repeated in the \yFilter\ band for samples of $z\sim5$ and $z\sim6$ galaxies, extracted from the optical HUDF catalogs. The \yFilter\ corresponds to 1750\AA\ and 1500\AA\ restframe at $z\sim5$ and $z\sim6$ respectively.  For the $z\sim4$ population the \iFilter\ band corresponds most closely to the 1600\AA\ wavelength. Therefore morphological k-correction effects are minimal.

A Sersic profile is fitted to these stacked images from all three redshift bins and both luminosity ranges. 
Through simulations, we have verified that the stacking procedure reproduces the average profiles of a galaxy population accurately \citep[see also][]{hathi08a}.
In order to compare the different profiles directly, we have convolved the best-fit models with the \jFilter\ band PSF and show the result in Figure \ref{fig:profiles}. The faint $z\sim7$ galaxies are only marginally resolved. The average profile is remarkably similar at all redshifts from $z\sim4$ to $z\sim7$, except for the appearance of extended wings at $z\sim4$ for all luminosities, and extended wings at $z\sim5$ for the more luminous galaxies.  But even the similar angular size indicates increasing physical sizes towards lower redshifts due to increasing angular diameter distances (30\% larger at $z\sim4$ vs. $z\sim7$).

The steepness of all these profiles suggests that star-formation is largely centrally concentrated in these galaxies and not widely distributed in a large, clumpy disk.

\subsection{SFR surface density}

\begin{figure}[htbp]
	\centering
		\includegraphics[width=\linewidth]{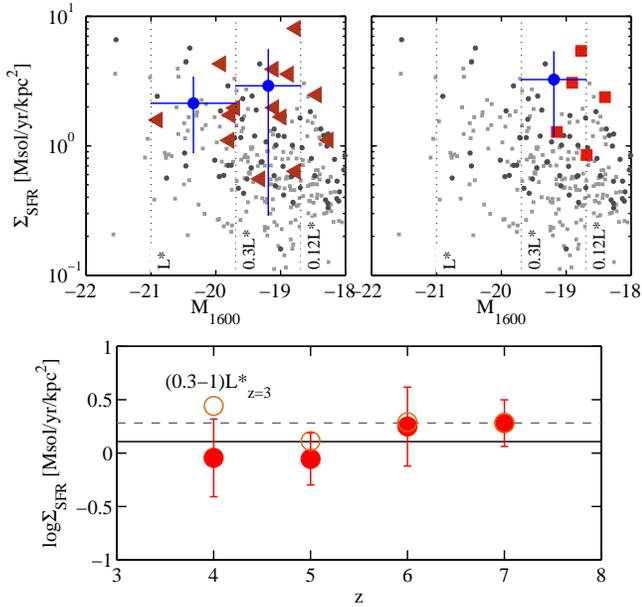}
	\caption{\textit{Top:}  The surface density of star-formation against luminosity. As in Figure \ref{fig:LumSizeSFR}, the gray background points represent the $z\sim4-6$ population in the WFC3/IR data and the colored points are the $z\sim7$ (left) and 8 population (right).
	\textit{Bottom: } The mean surface density of star-formation rate against redshift. Red dots correspond to the actual measurements, while the orange circles are corrected for dust absorption \citep{bouw09c}. The black line is the average for the uncorrected values and the dashed line the same for the dust corrected ones. In both cases, the average $\Sigma_{SFR}$ is consistent with being constant over the entire redshift range $z\sim4-7$.
	}
	\label{fig:sigmaSFRz}
\end{figure}

Given the slow evolution of the sizes of LBGs, it is interesting to estimate the change in the average surface density of star-formation, $\Sigma_{SFR}$.
The upper panel of Figure \ref{fig:sigmaSFRz} shows this as a function of absolute magnitude. The star-formation rate (SFR) has been estimated based on a simple conversion of the UV luminosity to SFR \citep{madau98} assuming a Salpeter initial-mass function. For the $z\sim7$ galaxies the SFRs range from $\sim1$ to $10M_\odot$yr$^{-1}$. Furthermore, our results suggest that the average $\Sigma_{SFR}$ remains relatively constant for the whole redshift range from $z\sim7$ to $z\sim4$. This is shown for galaxies with luminosities in the range (0.3-1)$L^*_{z=3}$ in the lower panel of Figure \ref{fig:sigmaSFRz}.
Filled red circles represent the actual measured values, while the orange open circles are corrected for dust extinction using the formula by \citet{meurer99} and UV continuum slopes measured in this same luminosity and redshift ranges from \citet{bouw09c}. In agreement with the size evolution, there is a small trend towards lower apparent $\Sigma_{SFR}$ at lower redshifts. However, when corrected for dust extinction, the star-formation surface density is consistent with being constant over the entire redshift range $z\sim3-7$.

A possible explanation for the constancy of the star-formation surface density is that the average star-formation efficiency is very similar in all these galaxies and that feedback effects change the mode of star-formation only little in these primordial galaxies. Note that this result is consistent with a constant limiting surface brightness from $z\sim7$ down to local galaxies \citep[see][]{meurer97,hathi08b}.
However, there are exceptions from this typical mode of star-formation. For example, the average surface density of star-formation is almost three orders of magnitudes larger in hyper-starbursts in quasar hosts at similar redshifts \citep[$z=6.42$;][]{walter09}.

\section{Summary and Conclusions}
\label{sec:summary}
We have used the ultra-deep WFC3/IR observations of the HUDF09 program to study the structures and morphologies of $z\gtrsim6.5$ galaxies previously presented in \citet{oesch09b} and \citet{bouw09b}.
These galaxies are all extremely compact, with an average size of $0.7\pm0.3$ kpc. Only two out of the 16 $z\sim7$ galaxies show a double core.  This fraction is slightly smaller than the $\sim30\%$ of LBGs with disturbed morphologies and double cores found at lower redshifts \citep[e.g.][]{lotz06,conselice09,petty09}, although they are still consistent within the small number statistics.
The $z\sim7$ galaxies are extremely similar to the $z\sim6$ population, both in sizes and in their average light profiles, showing that galaxy evolution proceeds slowly over the $\sim170$ Myr from $z\sim7-6$.

By comparison to LBGs down to $z\sim4$ only a very slow size evolution is found, following $(1+z)^{-m}$, with $m=1.12\pm0.17$ for galaxies of luminosities (0.3-1)$L^*_{z=3}$. Fainter galaxies down to $0.12L^*_{z=3}$ follow a similar scaling with $m=1.32\pm0.52$.
Additionally, the mean star-formation surface densities of LBGs are found to be constant over the entire redshift range $z\sim4-7$, which may be explained by largely constant star-formation efficiencies at these early epochs.

Similar exponents for the size scalings with redshift have been found for disk galaxies between $z\sim0-3$ \citep[e.g.][]{buitrago08}, as well as from semi-analytical modeling including concentrated dark-matter halo profiles \citep[e.g.][]{somerville08,firmani09}

By extending the present study to larger samples of $z\sim7$ galaxies which will become available in the near future it will be possible to put more stringent constraints on galaxy evolution models and to shed more light on the question of the main driver of star-formation in LBGs at these early epochs.

\acknowledgments{We especially thank all those at NASA, STScI and throughout the community
who have worked so diligently to make Hubble the remarkable observatory
that it is today. The servicing missions have
rejuvenated HST and made it an extraordinarily productive scientiﬁc
facility time and time again, and we greatly appreciate the support of
policymakers, and all those in the flight and servicing programs who
contributed to the repeated successes of the HST servicing missions.
PO acknowledges support from the Swiss National Foundation (SNF). 
This work has been supported by NASA grants NAG5-7697 and
HST-GO-11563.01.
}

Facilities: \facility{HST(ACS/NICMOS/WFC3)}.

\end{document}